\documentclass[
 aip,
 amsmath,amssymb,
 reprint,
 floatfix
]{revtex4-2}
\usepackage{siunitx}
\usepackage{graphicx}
\usepackage{dcolumn}
\usepackage{bm}

\usepackage[utf8]{inputenc}
\usepackage[T1]{fontenc}
\usepackage{mathptmx}
\usepackage{etoolbox}
\usepackage{textgreek}

\makeatletter
\def\@email#1#2{
 \endgroup
 \patchcmd{\titleblock@produce}
  {\frontmatter@RRAPformat}
  {\frontmatter@RRAPformat{\produce@RRAP{*#1\href{mailto:#2}{#2}}}\frontmatter@RRAPformat}
  {}{}
}
\makeatother
\begin{document}

\preprint{AIP/123-QED}

\title[Nonequilibrium Dynamics of the Helix-Coil Transition]{Nonequilibrium Dynamics of the Helix-Coil Transition in Polyalanine}

\author{Maximilian Conradi}
\email{maximilian.conradi@itp.uni-leipzig.de}
\affiliation{ 
  Institut für Theoretische Physik, Universität Leipzig, IPF 231101, 04081 Leipzig, Germany
}
\author{Henrik Christiansen}
\affiliation{
  Institut für Theoretische Physik, Universität Leipzig, IPF 231101, 04081 Leipzig, Germany
}
\affiliation{NEC Laboratories Europe GmbH, Kurfürsten-Anlage 36, 69115 Heidelberg, Germany}
\author{Suman Majumder}
\affiliation{
Amity Institute of Applied Sciences, Amity University Uttar Pradesh, Noida 201313, India
}
\author{Fabio Müller}
\affiliation{ 
  Institut für Theoretische Physik, Universität Leipzig, IPF 231101, 04081 Leipzig, Germany
}
\author{Wolfhard Janke}
\affiliation{ 
  Institut für Theoretische Physik, Universität Leipzig, IPF 231101, 04081 Leipzig, Germany
}
\date{\today}

\begin{abstract}
  In this work, the nonequilibrium pathways of the collapse of the helix-forming biopolymer polyalanine are investigated. To this end, the full time 
  evolution of the helix-coil transition is simulated using molecular dynamics simulations. At the start of the transition short $3_{10}$-helices form,
  seemingly leading to the molecule becoming more aspherical midway through the collapse. After the completed collapse, the formation of $\alpha$-helices
  seems to become the prevalent ordering mechanism leading to helical bundles, a structure representative for the equilibrium behavior of longer chains.
  The dynamics of this transition is explored in terms of the power-law scaling of two associated relaxation times as a function of the chain length.  
\end{abstract}

\maketitle

\section{Introduction}

The shape of a protein can influence its function or lead to a loss of function altogether, which is the cause of various diseases.\cite{dob} 
While machine learning based approaches are able to accurately predict the structure of a protein based on its sequence of amino acids,\cite{AlphaFold,CASP}
we still do not understand the underlying mechanisms of secondary and tertiary structure formation. Specifically, the dynamics of the nonequilibrium 
pathways that enable proteins to fold into their native states in time scales of micro- to milliseconds is not well understood.\cite{dil:ozk,dil:mac} In
contrast, nonbiological homopolymers like polystyrene take multiple seconds to fold.\cite{chu:yin} 

The collapse of the protein backbone plays a major role in this process. The impact on the folding originates from backbone-angle 
preferences and the chain entropy, which influence the native state of the protein.\cite{dil:mac,red:thi,har2012,sad:lap:mun,cam:thi} Based on this hypothesis numerous
studies have investigated this process using homopolymers and polypeptides as model for the backbone.\cite{deg,hal:gol,byr:kie,tim:kuz:yu,Kuz:tim1995,Kuz:tim1996,abr:lee:obu,Kik:ryd2005,xu:zhu,Guo:lia:wan,udg,maj:jan2015,maj:jan,maj:jan2016,maj:zie:jan,chr:maj:jan,maj:han:jan,maj:chr:jan} 
Only in recent decades have experimental methods like small-angle X-ray scattering, single-molecule fluorescence or dielectric spectroscopy been able to monitor the collapse of a single molecule.\cite{eat:mun,hag:eat,chu:hsi,don2001,kub:hof:eat,haa2005,rod:mak,ara:kay:mat,kon:kim:mat,kat:guo:gra,har2012,ban:den}
It is for this reason that most earlier studies used computational methods like Monte Carlo (MC) simulations or molecular dynamics (MD) simulations to analyze 
this transition. In this regard, MD simulations have been used more extensively not only because of easily available open source frameworks, but also due 
to the fact that the physical dynamics can be represented in MD simulations more faithfully than in MC simulations.

In general, a polymer undergoes a transition from an extended, random coil to an ordered, compact state, when the solvent condition is changed from good 
to poor. The first theoretical description of the kinetics of this process has been proposed by de Gennes.\cite{deg} In his seminal ``sausage model'' the
collapse is explained by the formation of sausage-like intermediate structures due to interactions between monomers. This intermediate slowly becomes 
more spherical as the surface energy decreases until it finally reaches a globular state. Later Halperin and Goldbart provided an alternative 
phenomenological ``pearl-necklace'' theory.\cite{hal:gol} There, the collapse is described as a stepwise sequence of specific events during the
collapse. In the first stage, small clusters start to form along the chain. This leads to a sequence of interconnected pearls on the chain, the 
pearl-necklace intermediate. The pearls then start to slowly absorb monomers of the connecting segments. Eventually, the clusters merge together and form
a single globular structure. This theory is supported by numerous studies of collapse transitions and hence has become the commonly accepted picture for 
this type of process.\cite{byr:kie,tim:kuz:yu,Kuz:tim1995,Kuz:tim1996,chr:maj:jan,maj:zie:jan,maj:chr:jan,maj:jan2015,maj:jan2016,maj:han:jan} However, 
a newer study suggests that depending on the solvent viscosity and temperature a combination of both the pearl-necklace and sausage scenario can be 
observed.\cite{maj:chr:jan2024} While these theories do describe the collapse of a coarse-grained homopolymer, it is unclear whether this translates to biological
homopolymers and proteins.  

A quantity of significant interest in such studies of the nonequilibrium dynamics is the behavior of the collapse time $\tau_{\rm c}$ as a function of the 
length of the polymer (typically measured by the number of monomers or residues), which usually follows a power-law scaling of the form
\begin{equation}
  \tau_{\text{c}}\sim N^z,
  \label{tau}
\end{equation}
with $z$ being the dynamic exponent. Past studies have found different values for the exponent z, ranging from $z\approx1-2$, depending
on the specifics of the simulation. Generally, MC simulations\cite{hal:gol,byr:kie,tim:kuz:yu,Kuz:tim1995,Kuz:tim1996} provide higher values of 
$z \approx 2$ compared to MD simulations\cite{abr:lee:obu,Kik:ryd2005,Guo:lia:wan} where $z \approx 1$ is found; for an overview, see Ref.\ 28. 
This is explained by the fact, that MD simulations are able to incorporate hydrodynamic effects resulting in a faster dynamics. 

All these results have been obtained from simulations of coarse-grained homopolymers, which 
typically represent non-biopolymers. Similar studies considering biopolymers are rare. 
Only recently, an all-atom MD simulation study of the collapse dynamics of polyglycine, 
a bio-homopolymer composed of the amino acid glycine as its residue, in explicit water has 
been reported.\cite{maj:han:jan} There, although the sequence of events during the collapse is similar 
to that of non-biopolymers, the dynamics is found to be even faster with $z = 1/2$. The authors 
argued that this is caused by the instantaneous hydrogen-bond formation, and speculated 
that the faster collapse of protein molecules can be attributed to this ultra-fast
dynamics of their polypeptide backbone. However, glycine does not feature any secondary structure formation. Hence, the
effect of secondary structure formation on the collapse dynamics is still unexplored. In 
this context, one of the simplest polypeptides that shows a distinct secondary structure formation, 
i.e., helix formation, is polyalanine.\cite{pal:ble2011} Alanine contains a methyl group CH$_3$ attached to
the --CONH$_2$ group, in contrast to the H-atom in glycine. The presence of this bulky methyl 
group increases its helical propensity resulting in the formation of a helical conformation 
at low temperature, thus exhibiting a helix-coil transition as a function of temperature.\cite{ara:han,ara:han2007}

Thus, with the motivation to understand the effect of secondary structure formation on the 
overall collapse of a polypeptide, here, we explore the relaxation dynamics of the collapse 
of the helix forming biopolymer polyalanine by means of all-atom MD simulations in implicit 
solvent. Our results reveal that due to the tug-of-war between the collapse and secondary 
structure formation, the sequence of events observed in the nonequilibrium pathway is 
significantly different from what is observed for a non-biopolymer, unlike the collapse of 
polyglycine. We show that in addition to the traditional methods of extracting relaxation 
times from the time dependence of the radius of gyration, the behavior of various shape 
factors also allows us to extract related relaxation times. To quantify our observations, 
we also investigate the scaling of these relaxation times with the length of the polypeptide 
which in all cases shows a much slower dynamics than what is observed for polyglycine, 
suggesting a slowing down of the overall collapse dynamics due to the simultaneous formation 
of secondary structure.
 
The rest of the paper is structured as follows. In Sec.\ II, we describe the model and 
details of our simulation method along with the definition of various observables. In 
Sec.\ III, we present our results which contains a qualitative picture of the 
transition, a quantitative description of the same, and analyses of scaling properties of
various relaxation times. Finally, in Sec.\ IV, we provide a short summary of our 
conclusions as well as an outlook on future research options.

\section{Model and Methods}
\label{mome}
Molecules of (Ala)$_{N}$ are prepared with a hydrogenated N-terminus (-NH$_2$) and a C-terminus (-COOH).
The all-atom MD simulations are run using the OpenMM package.\cite{openMM} The Amber14 force field\cite{mai:mar} is employed for interactions between the
atoms. All simulations are performed in a generalized Born implicit solvent model.\cite{ngu:roe} For the nonbonded interactions a cutoff radius of $1$ nm
and a switch-off distance of $0.9$ nm are used. Initially, the molecules are equilibrated at $T=2000$ K for at least $1$ ns, where they take random-coil conformations.
Following this, the molecules are quenched to a temperature of $T_{q} = 300$ K significantly below the transition temperature $T_{\text{c}}$, 
which is at least $415$ K for a short chain of length $N=20$ residues.\cite{ara:han,ara:han2007} MD simulations are performed using the Langevin 
thermostat with a leap-frog integration scheme. We choose a step size of $\delta t = 1$ fs and a friction coefficient of $\gamma = 1$ ps$^{-1}$ for the
integrator. We have simulated polyalanine molecules with twelve different chain lengths $N=25, 50, 75, \dots, 300$ residues. For chain lengths up to $N=100$ we use 200 independent 
initial conformations and for longer chains 100 initial conformations each. Each simulation is run for at least $100$ ns and measurements are performed in 
logarithmically spaced intervals. The results are averaged over 
all realizations for each considered chain length $N$. Results for the relaxation times are obtained using the delete-one Jackknife method.\cite{efr,efr:ste,mil}

The results of the simulations are analyzed using three main types of observables, which we introduce in the following: (1) Rotationally invariant quantities 
derived from the gyration tensor, (2) energy contributions based on the force field, and (3) hydrogen bonds and secondary structure. Based on the positions
of the individual atoms $\vec{r}_m$ at a given point in time the components of the gyration tensor $\bm{Q}$ can be calculated as follows:
\begin{equation}
  \bm{Q} = Q_{ij} = \frac{1}{M}\sum_{m=1}^M (r_m^i - r_{\text{CM}}^i)(r_m^j - r_{\text{CM}}^j),\qquad i,j = 1, \dots, d.
\end{equation}
Here, $M$ is the total number of atoms in the model, $r_{\text{CM}}^i$ refers to the $i$-th component of the center of mass vector, and $d=3$ is the 
spatial dimension. From this we can derive the squared radius of gyration $R_{\text{g}}^2$ as
\begin{equation}
  R_{\text{g}}^2 = \frac{1}{M}\sum_{m=1}^{M}{(\vec{r}_m - \vec{r}_{\text{CM}})}^2 = \sum_{i=1}^{d} {Q}_{ii} = \mathrm{Tr}\hspace*{1mm}\bm{Q}.
\end{equation}
Three more quantities that can be derived from the gyration tensor are the asphericity $A$, the prolateness $S$, and the nature of asphericity $\Sigma$. Using the average eigenvalue 
of the gyration tensor
\begin{equation}
  \bar{\lambda}=\frac{\mathrm{Tr}\hspace*{1mm}\bm{Q}}{3}=\frac{1}{3}\sum_{i=1}^{3}\lambda_i , 
\end{equation}
we can define the asphericity as\cite{the:sut,bla:jan,ark:jan}
\begin{equation}
  A = \frac{1}{6}\sum_{i=1}^{3} \frac{{(\lambda_i-\bar{\lambda})}^2}{\bar{\lambda}^2} = \frac{3}{2} \frac{\mathrm{Tr}\hspace*{1mm}\hat{\bm{Q}}^2}{{(\mathrm{Tr}\hspace*{1mm}\bm{Q})}^2},
\end{equation}
where $\hat{\bm{Q}} = \bm{Q} - \bar{\lambda}\bm{I}$ and $\bm{I}$ is the unit matrix. This quantity describes the shape of a polymer based on how spherical it is.
It takes a value between $A=0$ for a perfectly spherelike conformation, where $\lambda_i=\bar{\lambda}$, and $A=1$ for a completely straight rodlike conformation
with all but one eigenvalue equal to zero. The second quantity is 
the prolateness that is given by\cite{bla:jan,ark:jan}
\begin{equation}
  S = \frac{\prod_{i=1}^{3}(\lambda_i - \bar{\lambda})}{\bar{\lambda}^3} = 27 \frac{\mathrm{\det}\hspace*{1mm}\hat{\bm{Q}}}{{(\mathrm{Tr}\hspace*{1mm}\bm{Q})}^3}.
\end{equation}
For a fully prolate, rodlike conformation one obtains $S=2$ since $\lambda_1\not=0$ and $\lambda_2=\lambda_3=0$. For an absolutely oblate, disklike conformation,
on the other hand, one has $\lambda_1=\lambda_2$ and $\lambda_3=0$ which results in $S=-1/4$. This implies $-1/4\leqq S\leqq2$. Since $S=0$ for a spherelike conformation, in general $S>0$ corresponds to 
a prolate, ellipsoidlike conformation and $S<0$ to an oblate one. Finally, one can calculate the nature of asphericity\cite{ali:fre,ost:ali:fre} 
\begin{equation}
  \Sigma = \frac{4\prod_{i=1}^{3}(\lambda_i - \bar{\lambda})}{{\left[\frac{2}{3}\sum_{i=1}^{3}{(\lambda_i - \bar{\lambda})}^2\right]}^{3/2}} = \frac{4\mathrm{\det}\hspace*{1mm}\hat{\bm{Q}}}{{(\frac{2}{3}\mathrm{Tr}\hspace*{1mm}\hat{\bm{Q}}^2)}^{3/2}}.
\end{equation}
This quantity gives values of $\Sigma=-1$ for a disk and $\Sigma=1$ for a rigid rod. Hence, it has the boundaries $-1\leqq\Sigma\leqq1$. 
Note that for any given conformation, the latter three shape parameters are related by $\Sigma = S/(2A^{3/2})$.

Additionally to the aforementioned geometrical quantities, we also monitor the individual components of the energy as obtained from the Amber force field.\cite{mai:mar,cor:cie} 
The energies calculated by the force field can be divided into nonbonded energies, which are calculated for atoms that are three or more bonds apart, and
bonded energies. The first group consists of the energy based on Coulomb interactions between two charged particles and the energy derived from the 
Lennard-Jones interactions for pairs of atoms. In regard to the bonded energies we specifically look at the torsion energy, since it is known that the 
dihedral angles are related to the secondary structure. 

Finally, using the DSSP algorithm, one identifies hydrogen bonds and based on their pattern assigns secondary structures to residues.\cite{DSSP} 
This way we can determine residues that are part of helical segments or other secondary structure elements. One of the two most common secondary structure
elements are $\alpha$-helices which are present in most proteins.\cite{bar:tho} Its basic structure consists of backbone amide bonds between residues $i$
and $i+4$ marking a turn with a length of approximately $3.6$ residues and an offset of $0.54$ nm.\cite{mun} Additionally, there exist two other types of
helical structures: $3_{10}$- and $\pi$-helices. $3_{10}$-helices are helix structures that are based on bonds between residues $i$ and $i+3$ making them
narrower than $\alpha$-helices. Usually, this type is only found in short sequences of 3 to 4 residues at the ends of $\alpha$-helices.\cite{bar:tho,nem:phi,BAK:hub} Therefore, they are viewed
as a possible intermediate in the formation of $\alpha$-helices.\cite{mun} Finally, $\pi$-helices consist of bonds between residues $i$ and $i+5$. 
We did not encounter this helix type as it is relatively rare because the backbone angles are energetically unfavorable compared to $\alpha$-helices and 
the entropic cost for the formation is higher than for the other two helix types.\cite{car:doi} 

\section{Results}
\subsection{Qualitative picture of the helix-coil transition in polyalanine}
\begin{figure*}[t!]
\begin{tabular*}{\linewidth}{p{.33\linewidth}p{.33\linewidth}p{.33\linewidth}p{0pt}}
    \centering
    \includegraphics[width=.8\linewidth]{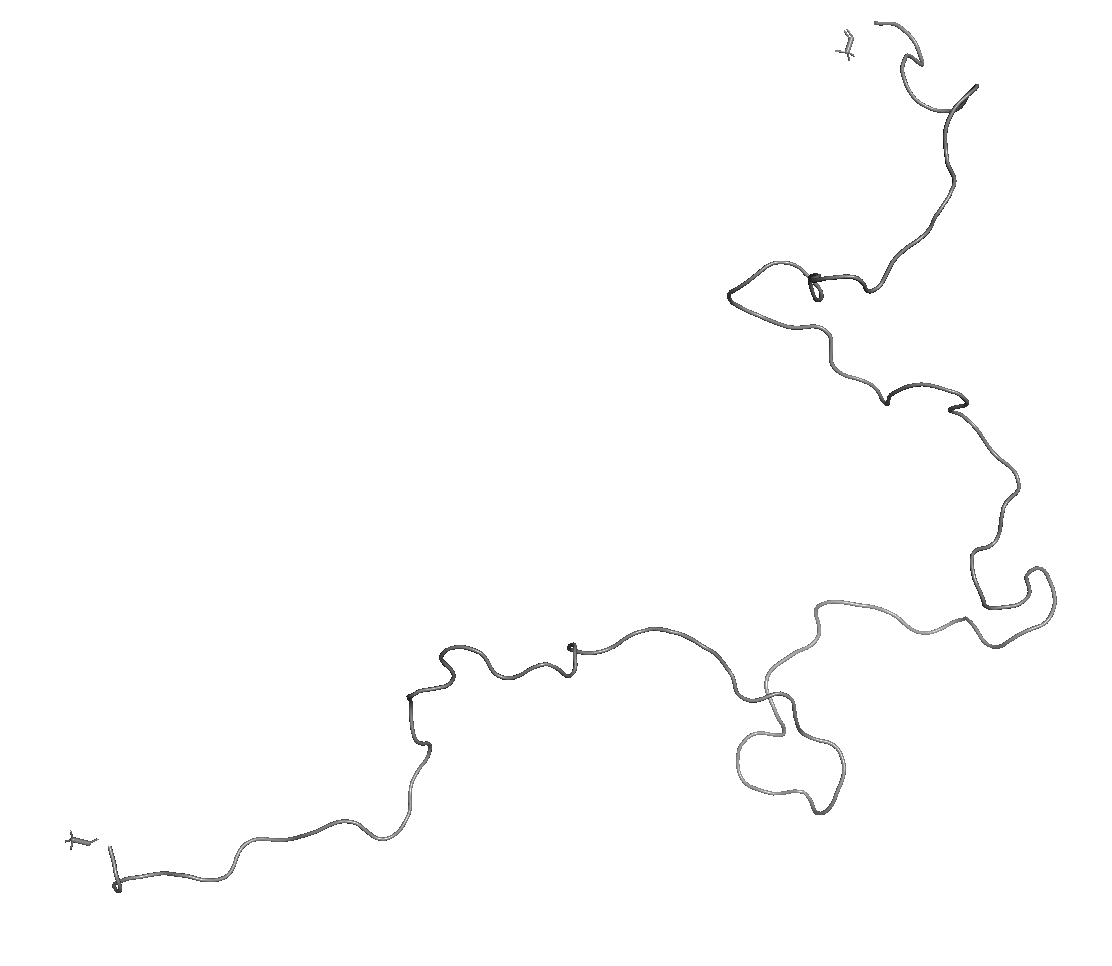}
    &
    \centering
    \includegraphics[width=.8\linewidth]{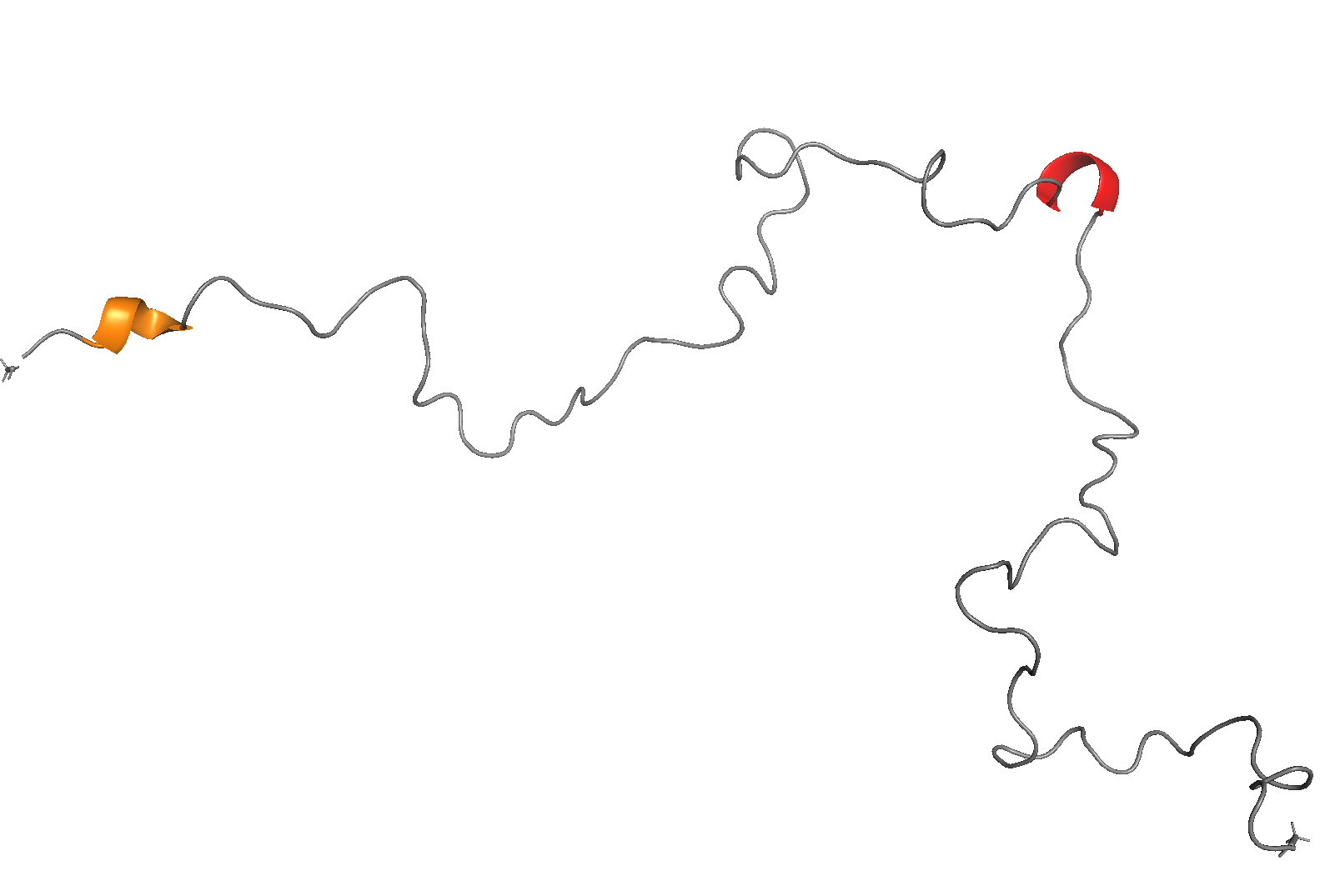}
    &
    \centering
    \includegraphics[width=.8\linewidth]{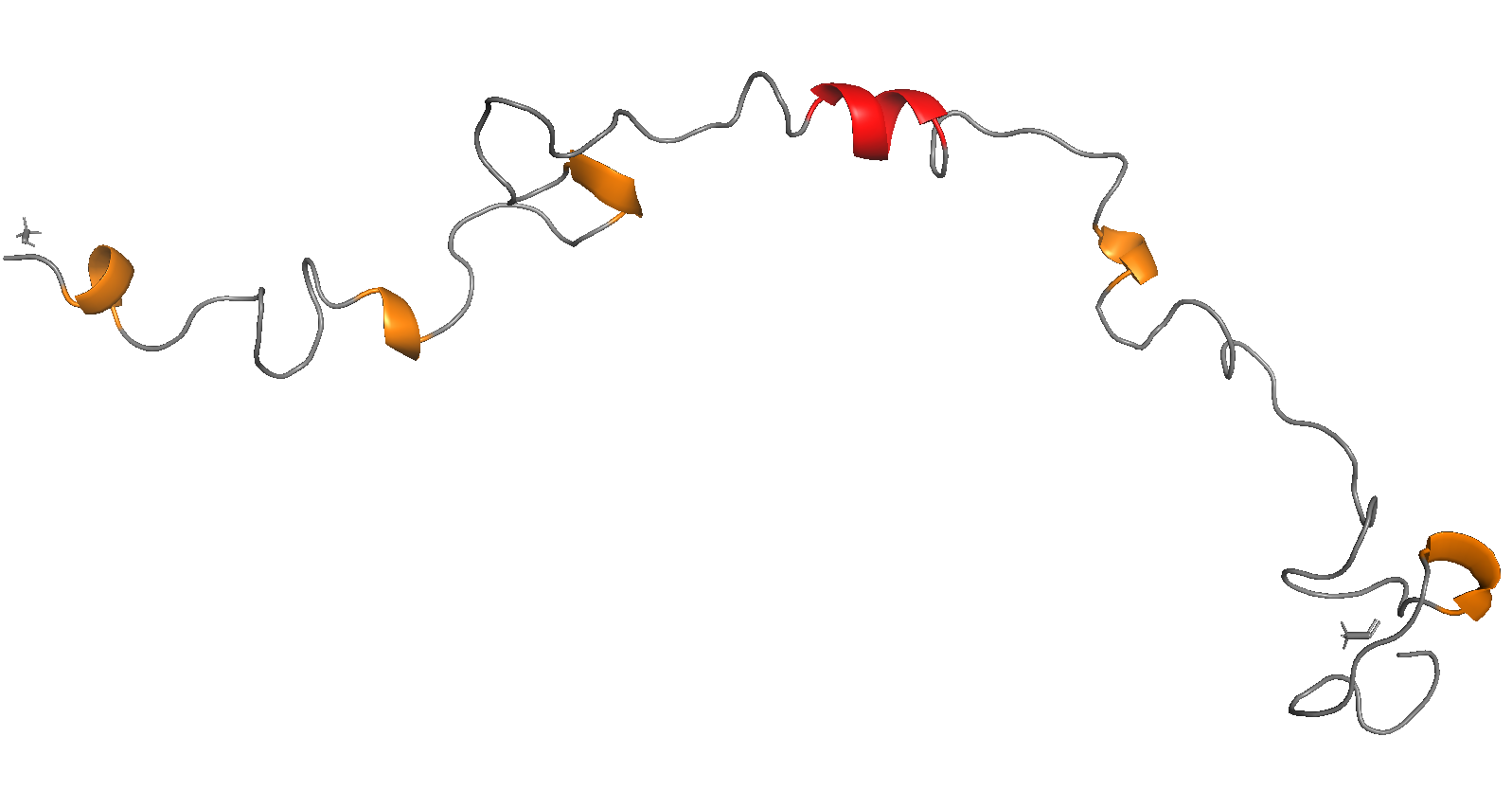}&
    \\
    \centering
    \includegraphics[width=\linewidth,page=1]{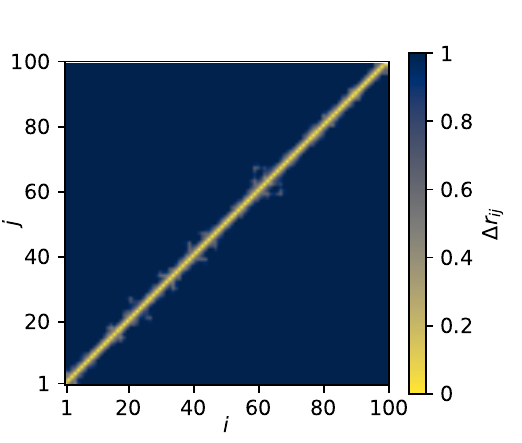}
    &
    \centering
    \includegraphics[width=\linewidth,page=2]{ContactMaps2.pdf}
    &
    \centering
    \includegraphics[width=\linewidth,page=3]{ContactMaps2.pdf}&
    \\
    \centering
    $t=\SI{0}{\nano\second}$
    &
    \centering
    $t=\SI{0.003}{\nano\second}\approx t_0$
    &
    \centering
    $t=\SI{0.15}{\nano\second} = t_1$&
    \\
    \centering
    \includegraphics[width=.6\linewidth]{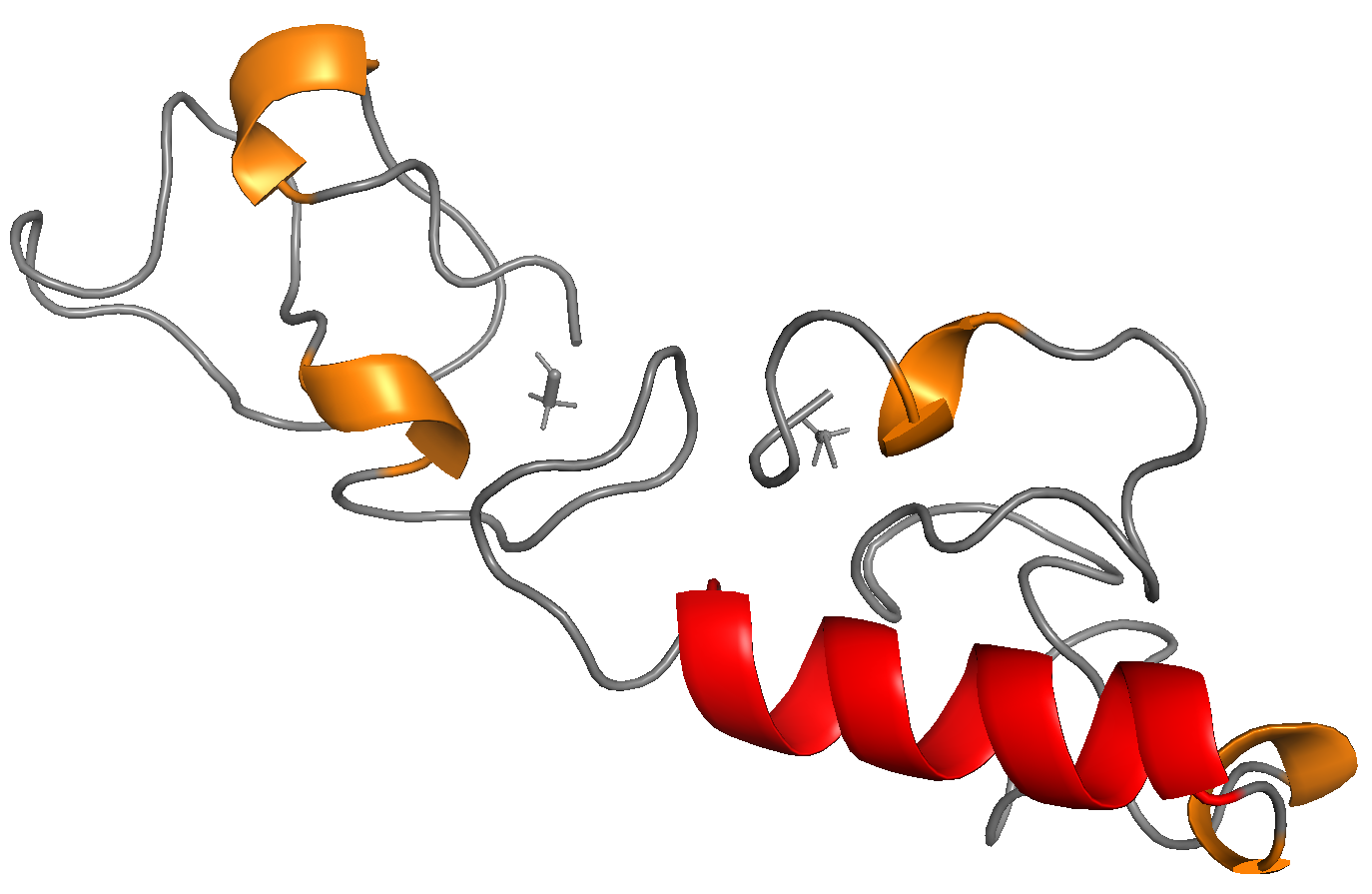}
    &
    \centering
    \includegraphics[width=.6\linewidth]{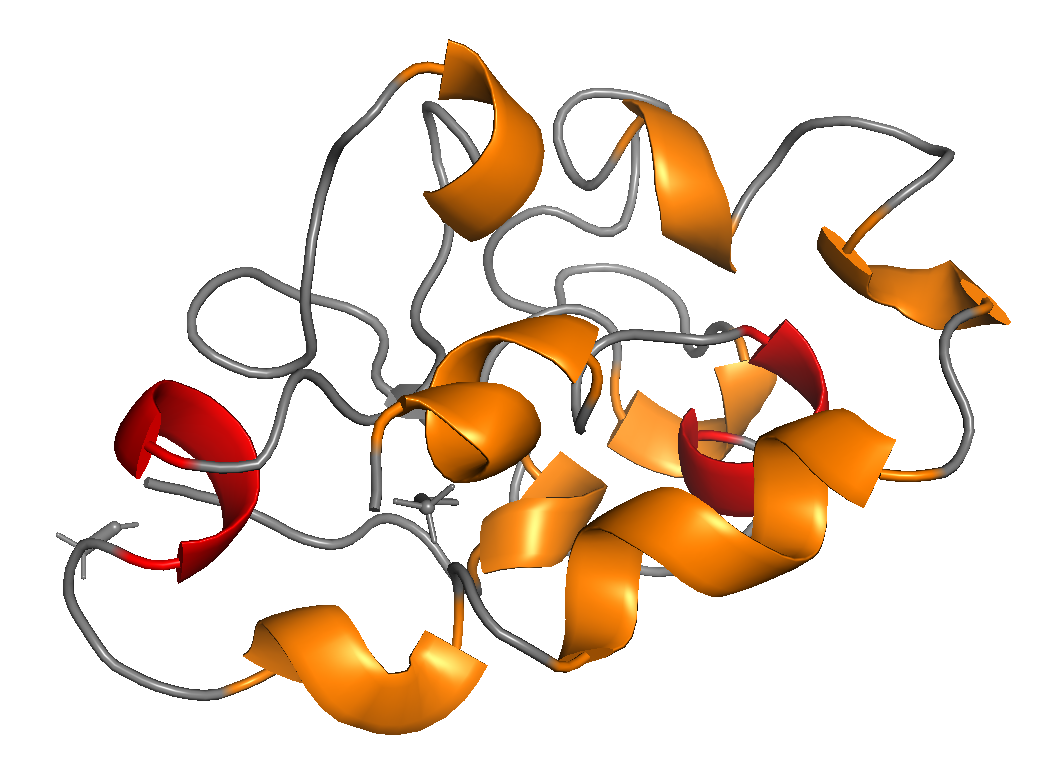}
    &
    \centering
    \includegraphics[width=.5\linewidth]{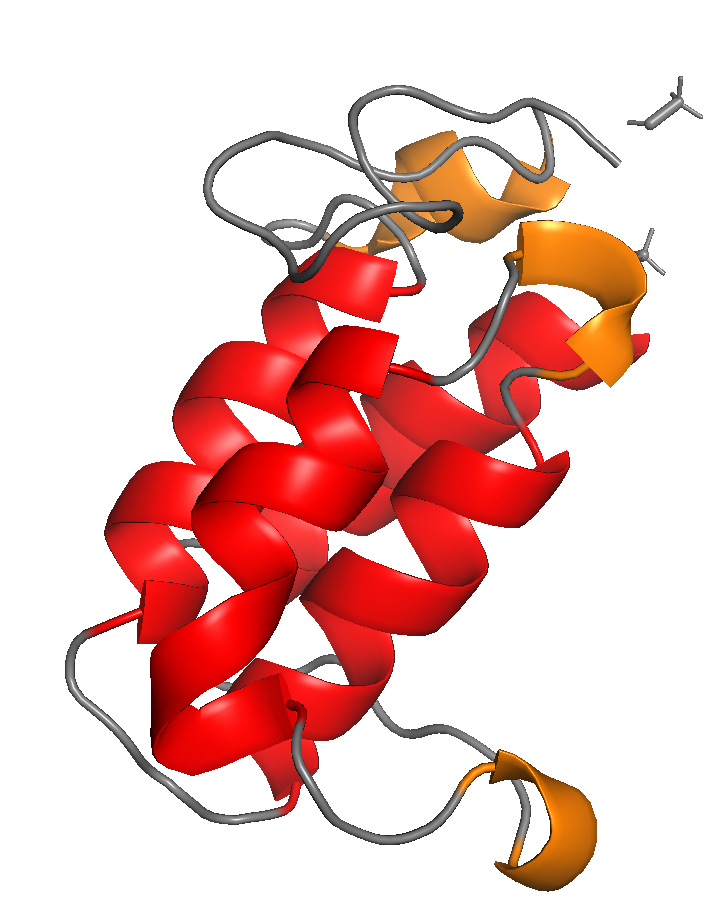}&
    \\
    \centering
    \includegraphics[width=\linewidth,page=5]{ContactMaps2.pdf}
    &
    \centering
    \includegraphics[width=\linewidth,page=6]{ContactMaps2.pdf}
    &
    \centering
    \includegraphics[width=\linewidth,page=8]{ContactMaps2.pdf}&
    \\
    \centering
    $t=\SI{1}{\nano\second}\approx t_2$
    &
    \centering
    $t=\SI{10}{\nano\second}$
    &
    \centering
    $t=\SI{1000}{\nano\second}$&
    \\
\end{tabular*}
\caption{Snapshots during the transition of a single polyalanine molecule with length $N=100$. $\alpha$-helical residues are colored red, 
$3_{10}$-helical residues are colored orange. Below each snapshot is the corresponding contact map. The distances $\Delta r_{ij}$ between residues $i$ and $j$ 
are given in nm. Distances above $\Delta r_{ij}>1$ nm are set to $1$ nm. Colors are assigned for each distance $\Delta r_{ij}$ based on the color bar on 
the right side of the map. The times $t_0, t_1, t_2$ mark specific events in the trajectory discussed in the following section.}\label{traj}
\end{figure*}

\begin{figure}[t!]
  \centering
  \includegraphics[width=\linewidth]{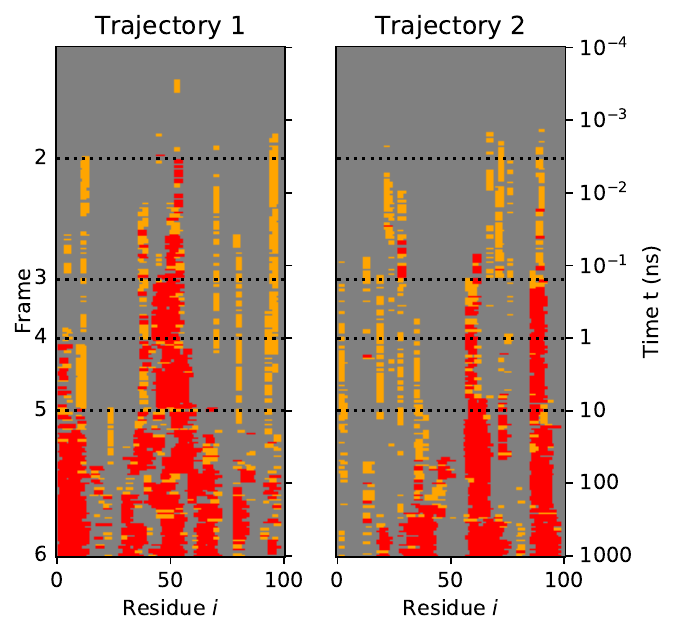}
  \caption{Time evolution of helices in two selected conformations. The left side shows the time evolution displayed in Fig.\ \ref{traj}, the right side
  another randomly chosen time evolution. $\alpha$-helical residues are colored red, $3_{10}$-helical residues are colored orange. The dotted lines mark 
  the times of the snapshots in Fig.\ \ref{traj}. Frame 1 is not shown.}\label{evo}
\end{figure}

We first qualitatively investigate the collapse kinetics of a single polyalanine molecule of length $N=100$ residues simulated up to $1000$ ns. In Fig.\ \ref{traj} snapshots from 
the trajectory of an individual molecule are displayed. Below each snapshot is the corresponding contact map. There, the distances $\Delta r_{ij}$ between the 
C$^{\alpha}$-atoms of residues $i$ and $j$ are displayed. The colors encode the distance, where yellow corresponds to small separations and
blue to larger values of $\Delta r_{ij}$. We apply a cutoff for distances $\Delta r_{ij}>1$ nm, to increase the visibility of contacts. 
The diagonal represents the contacts $\Delta r_{ii}=0$ nm and is hence colored in yellow.
In the first image one can see that before the quench the molecule is in an extended, random-coil state without helical residues. This is also
visible in the corresponding contact map where almost no yellow regions aside from the diagonal exist. 
Following the quench the molecule still remains in an extended conformation for some time which can be seen in the second frame at $t=0.003$ ns. Nonetheless,
first helical segments have formed along the chain as visible in the snapshot and in the contact map where they appear as yellow signals close to the diagonal.
In general, the molecule has slightly contracted in some areas.
At $t=0.15$ ns a short straight $\alpha$-helix with a length of two turns is present in the molecule. In the contact map this structure is indicated by the
broader yellowish region of close contacts around the diagonal approximately between residues 50 and 60. On both sides of this area, small yellow patterns
of close contacts appear on the diagonal, signaling the short $3_{10}$-helices that have formed. Off the diagonal there are now also some contacts between
residues farther apart. Overall, the molecule has straightened in comparison to the previous frames as can be seen from the snapshot. 
At $t=1$ ns the molecule takes a much more compact conformation where the segments on both parts of the helix are visibly stronger contracted. This
is also reflected in the contact map by yellow regions off the diagonal that signal contacts with residues that are significantly farther away. In addition,
the $\alpha$-helix has extended to more than three full turns and is visible as a broad yellow region on the diagonal. $3_{10}$-helices are also 
still present in the parts of the molecule on both sides of the $\alpha$-helix.
In the following, at $t=10$ ns, the residues in the molecule have rearranged to form a more compact globular structure in order to decrease the
surface exposed to the solvent. This is reflected in the contact map by close contacts even between residues that are very far apart such as $i=10-20$ 
and $j=90-100$. At this point, we also see in the snapshot that a number of short $3_{10}$-helices are present throughout the entire molecule. The longer 
$\alpha$-helix from the previous frame can still be seen in the bottom of the frame, although it is now classified as a $3_{10}$-helix.
Finally, in the last frame at $t=1000$ ns we see that the molecule has changed from a globular structure to a bundle of helices. The basis for this 
structure are the four $\alpha$-helices that are now aligned in the middle, each consisting of three to four turns. In the contact map this shows as broad
yellow stripes of close contacts that have formed alongside the diagonal. Aside from these regions next to the diagonal, we also observe 
yellowish patterns of close contacts between residues further apart. In contrast to previous frames, these brighter regions now appear in regular 
patterns of lines parallel and perpendicular to the diagonal. They belong to the contacts between two helices with perpendicular lines corresponding to 
antiparallel helices and parallel lines to parallel helices. This regular layout shows that the molecule is highly organized with the helices aligned 
into a bundle. The observed pattern already resembles contact maps for equilibrium conformations found in previous studies.\cite{pal:ble,palencar_folding_2013}

The evolution of the helices is even better displayed in Fig.\ \ref{evo}. Here, each residue is matched with the type of helix it is part of
at any point during the helix-coil transition. Trajectory 1 matches the conformations shown in Fig.\ \ref{traj}. In this frame multiple
vertical orange lines are visible. These represent $3_{10}$-helices that start to form early in the transition as shown in the snapshots. These structures
stay in place once they are formed resulting in line patterns. It is notable that all $3_{10}$-helices remain short, as visible in the frames of Fig.\ \ref{traj}.
The $\alpha$-helix in the middle of the conformation forms early in the transition in a spot where previously a $3_{10}$-helix existed. 
In the following this $\alpha$-helix expands in both directions by merging with adjacent $3_{10}$-helices and eventually bends into multiple helices as 
can be seen close to $t=10$ and $t=100$ ns. In the final part at $t=1000$ ns we can see four $\alpha$-helices, which are all of similar length. Notably, three
of them are only separated by a few residues and have seemingly emerged from the same $\alpha$-helix that formed early in this trajectory. Aside from these
a significant part of the molecule has no $\alpha$-helix structures, although these regions exhibited helices at earlier times. For comparison
the right frame in Fig.\ \ref{evo} displays another randomly chosen trajectory. Here, it also seems like the helices persist in the spot in which they were 
formed and that almost all the $\alpha$-helices originate from $3_{10}$-helices. The trajectories in Fig.\ \ref{evo}, however, also show some limitations
of this analysis with the DSSP algorithm where the classification of individual residues can oscillate between $3_{10}$-helix and $\alpha$-helix.

In general, we observe that already early on helical segments are forming throughout the molecule, with the majority being $3_{10}$-helices. The molecule
then seems to collapse in some regions while simultaneously straightening around the helical segments. Subsequently, the molecule collapses into
a single globular state, which already contains short $\alpha$- and $3_{10}$-helices. These $\alpha$-helices appear to form in spots that previously 
showed $3_{10}$-helices supporting observations that $3_{10}$-helices can act as an intermediate in the formation of $\alpha$-helices.\cite{mun} 
Following this, the $\alpha$-helices begin to extend which leads to a transition into a bundle of helices. This seems to deviate from 
the traditional pearl-necklace picture which is likely due to the formation of short helices early in the collapse and their specific bonding pattern. 
While these exemplary trajectories provide a good illustration of the path that the molecule follows towards equilibrium, they are not necessarily 
statistically representative and are thus backed up by a more quantitative analysis in the next section.  

\subsection{Quantitative picture of the helix-coil transition}
\begin{figure}[t!]
  \centering
  \includegraphics[width=\linewidth]{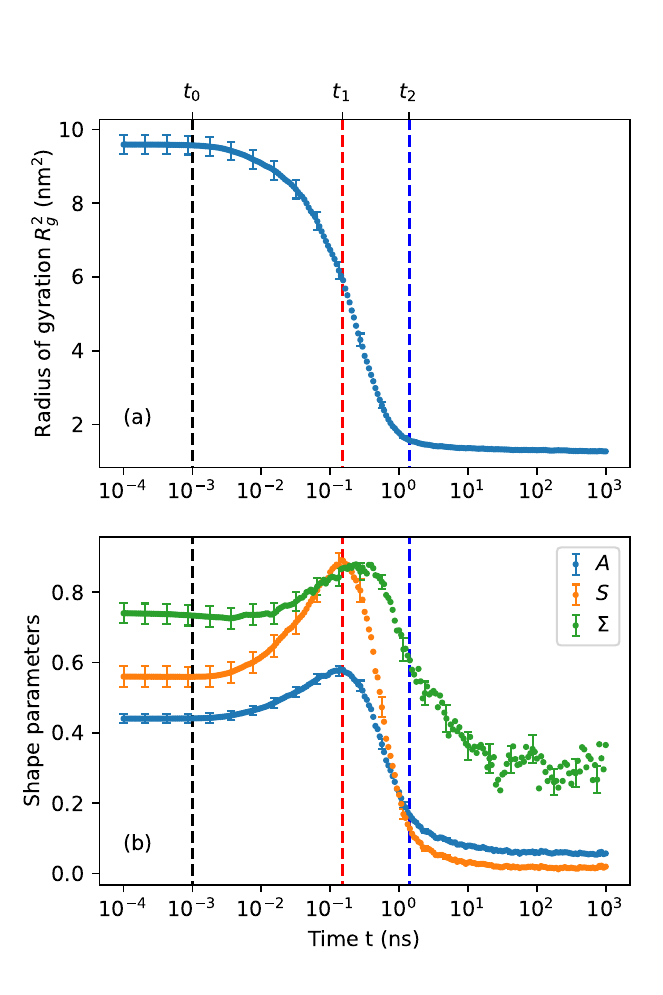}
  \caption{Time evolution of (a) the squared radius of gyration $R_g^2$ and 
  (b) asphericity $A$, prolateness $S$, and nature of asphericity $\Sigma$ for molecules of chain length $N=100$. The dashed vertical lines at $t_0 = 0.001$ 
  ns, $t_1 = 0.15$ ns, and $t_2 = 1.4$ ns mark notable events in the collapse. All results are averaged over 200 simulations with independent time 
  trajectories.}\label{force}
\end{figure}

Next, we consider the four shape parameters defined in Sec.\ \ref{mome}. For this we performed additional simulations with independent initial 
conformations for a total of 200 independent time trajectories. From the average of their time evolutions displayed in Fig.\ \ref{force} we identify 
three characteristic times $t_0, t_1$, and $t_2$. The times $t_0=0.001$ ns and $t_2=1.4$ ns are derived from the time dependence of the squared radius of
gyration $R_{\text{g}}^2(t)$ in (a), marking the beginning and end of the collapse, respectively. The time $t_1=0.15$ ns marks the maximum in asphericity 
$A$ and prolateness $S$ which is visible in (b). 
Looking at the time evolution of the squared radius of gyration we see that after the quench, the system needs a short amount of time to become unstable 
to fluctuations, i.e., to break the already-developed equilibrium correlations in the extended state at high $T$. This gets reflected in the behavior of 
$R_{\rm g}^2$ which initially is almost flat until $t_0$. From there it starts to decrease until the time $t_2$ is reached, where it stabilizes at a low 
value corresponding to the compact, globular state observed in Fig.\ \ref{traj}.

Comparing this to the other three shape parameters, we can once again see that initially after the quench all three quantities remain at a constant level 
taking on values that correspond to a random coil.\cite{bla:jan} Approximately at time $t_0$, when the collapse begins, both asphericity and prolateness
start to increase until they reach a maximum at $t_1$. From this point on, both observables show a rapid decrease until they reach stable values at 
approximately time $t_2$ once the molecule has fully collapsed. At this point, both quantities have decreased to values close to zero indicating that the
shape of the molecule has become almost spherical.  
In contrast, the nature of asphericity $\Sigma$ shows a slightly different time evolution. Similar to asphericity and prolateness, it initially 
remains at a stable value indicating no changes. This region of constant value continues past the beginning of the collapse at $t_0$ until approximately
$t\approx10^{-2}$ ns. Following this, $\Sigma$ starts to increase, albeit slower than the other two quantities. Consequently, it also reaches its maximum,
which indicates a rodlike shape, slightly later. Similar to the asphericity and prolateness it decreases afterwards, before fluctuating around a somewhat stable 
value of $\Sigma\approx0.3$ that indicates a slightly rodlike conformation. Strikingly, the maximum in the shape parameters happens in the middle of the collapse. This shows that despite the molecule collapsing, a more rodlike 
structure is able to form. 
A similar pattern of the time evolution for these shape parameters has also been observed for a non-biopolymer in a recent study.\cite{maj:chr:jan2024} 
There it was ascribed to a collapse separated into an initial pearl-necklace stage that results in a single sausage-like globule at the maximum of 
both asphericity and prolateness. This is then followed by a rearrangement of the residues into a spherical shape resulting in a decrease in $A$ and $S$ 
and, eventually, reaching a stable value. While we observe the same pattern in the time evolution of asphericity and prolateness, the structural 
properties reflected in the shape parameters are quite different. 
In fact, it is likely that in our case helical structures are related to this maximum in the three shape parameters. When comparing with the evolution of 
helices in Fig.\ \ref{evo} we can see that already at $t_0$ helices start to form in the molecule and at $t_1$ there are multiple short $3_{10}$-helices 
and one long $\alpha$-helix present in the molecule.

\begin{figure}[t!]
    \centering
    \includegraphics[width=\linewidth]{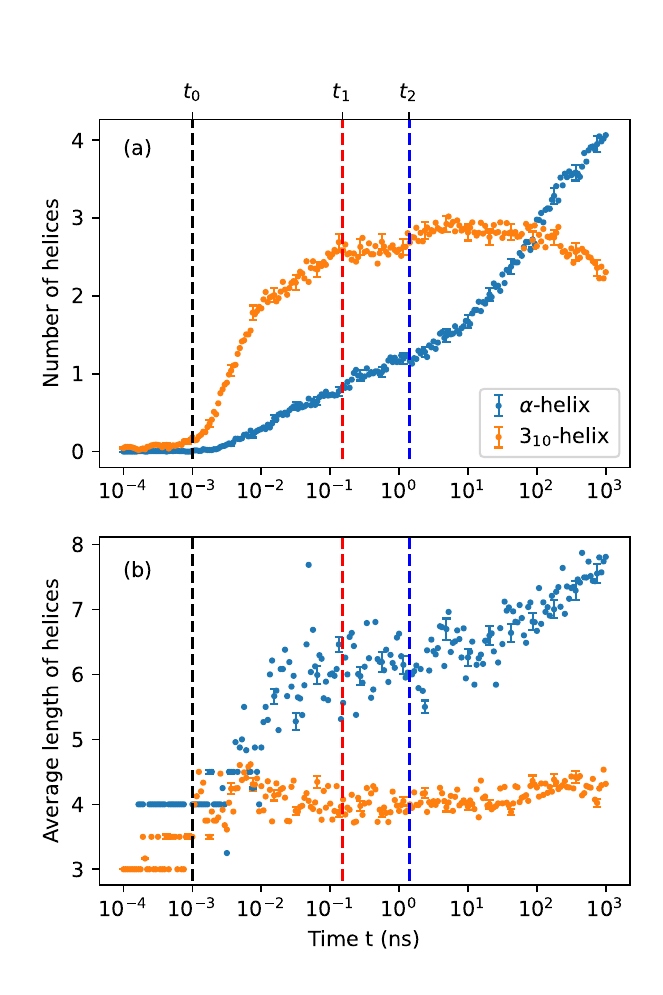} 
  \caption{Time evolution of (a) the average number of helices and (b) the average length of $\alpha$- and $3_{10}$-helices for molecules of chain length $N=100$.
  All results are averaged over 200 simulations with independent initial conformations. Legend for both frames is displayed in (a).}\label{helicity}
\end{figure}

The early formation of helices becomes even more apparent when looking at the time evolution of the average number and average length of helices in 
Fig.\ \ref{helicity}. Here, we can see in (a) that already with the beginning of the collapse at $t_{0}$, $3_{10}$-helices start to form as the number of 
helices increases. Shortly after the beginning of the collapse also $\alpha$-helices start to form. While the average number of $\alpha$-helices is 
growing throughout the entire transition, the number of $3_{10}$-helices starts to plateau at $t_1$, the maximum in asphericity and prolateness. This 
suggests that the formation of $3_{10}$-helices is related to the increase in asphericity and prolateness. After this point the average number of 
$3_{10}$-helices remains at about three helices per molecule. Only on longer time scales past the end of the collapse the number of $3_{10}$-helices is 
starting to slightly decrease again.

The average length of the helices in Fig.\ \ref{helicity} (b) shows a slightly different picture. Initially, the few existing $3_{10}$- and 
$\alpha$-helices show average lengths of approximately 3 and 4 residues, respectively, which is in both cases the minimum number of residues needed to form 
one turn. Close to $t_0$ the average length of the $3_{10}$-helices increases to approximately 4 residues. However, already before this point some 
$3_{10}$-helices were longer than 3 residues as indicated by the data points at an average length of 3.5 residues. The length of these helices then does 
not change throughout the remaining parts of the transition in accordance with previous observations that this helix type is only forming short helices.\cite{bar:tho,nem:phi,BAK:hub} 
The average length of $\alpha$-helices remains at the initial value longer and only increases when the number of $\alpha$-helices in Fig.\ \ref{helicity} (a) starts to 
increase. It then appears to increase until approximately $t_1$ where the average length fluctuates around a value of approximately 6 residues per helix.
The large fluctuations in the average length could be caused by the inaccuracies in the classification of DSSP that was already observed in Fig.\ \ref{evo}. 
After the end of the collapse the average length of $\alpha$-helices starts to increase again around the time at which also the number of 
$\alpha$-helices is showing a signal. This shows that, while able to form throughout the entire collapse, $\alpha$-helices are not able to extend between 
$t_1$ and $t_2$ where the molecule collapses to a more compact, globular state. Only at later times do the helices continue to extend while the molecule 
is transitioning into equilibrium states with longer straight $\alpha$-helices such as the helical bundle shown in the last frame of Fig.\ \ref{traj}. At
the end of our simulations $\alpha$-helices have reached an average length of 8 residues which is equivalent to two full turns. 
The average length at this time is close to what we observe in our own equilibrium simulations using precisely the same model setup, but significantly lower than what has been observed in vacuum where such
helices are more than double in size.\cite{pal:ble,pal:ble2011,palencar_folding_2013} The authors of these studies, however, also find a lower number of 
helices for the respective chain length. The reason for these longer helices, is that in the absence of a solvent extended conformations with long 
helices are stabilized.  

\begin{figure}[t!]
  \centering
  \includegraphics[width=\linewidth]{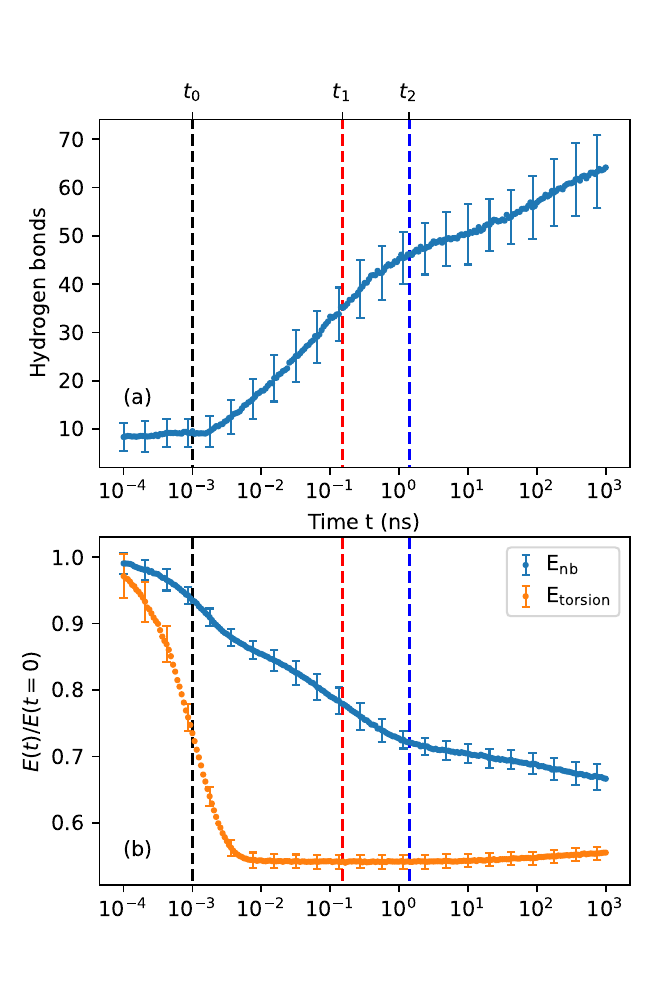}
  \caption{(a) Time evolution of the number of intramolecular hydrogen bonds. (b) Time evolution of the nonbonded energies $E_{{\text{nb}}}$ and torsion 
  energy $E_{\text{torsion}}$ normalized with their respective values at $t=\SI{0}{\nano\second}$. All results are for $N=100$ and averaged over 200
  simulations with independent initial conformations.}\label{helix}
\end{figure}

This pattern of helix formation is also displayed in the number of intramolecular hydrogen bonds in Fig.\ \ref{helix} (a). We can see that the number of 
hydrogen bonds is strongly correlated to the formation of helices, as one would expect. It starts to increase approximately at $t_0$ when $3_{10}$-helices
start to form and continues even after the end of the collapse, when only $\alpha$-helix-formation is happening.
Also in (b) the different contributions of the energy provided by the Amber14 force field\cite{mai:mar} display signals of helix formation.
The nonbonded energies start to decrease already early on, significantly before $t_0$. A first signal can be seen around
$t\approx10^{-2}$ ns, the time at which $\alpha$-helix formation is setting in. Further signals are observed at $t_1$ and around $t_2$ aligning with the 
peak in asphericity and prolateness and the end of the collapse. These signals in the time evolution could be related to the formation of helices and the difference in nonbonded energy between $3_{10}$- 
and $\alpha$-helices.\cite{tirado-rives_molecular_1993} 
The torsion energy on the other hand shows a simpler profile. Again, we observe that it decreases early on continuing past $t_0$ and 
reaches a constant level around the time when $\alpha$-helices begin to form, suggesting that the decrease in torsion energy could be related to the 
delayed formation of $\alpha$-helices. At late times no significant changes in the torsion energy
are visible. While harmonic bond energy and angle energy were also determined, they are not displayed as their behavior shows no specific signals for the
collapse.

In general, we are able to distinguish two stages of the transition: The collapse and the subsequent folding to a helical bundle. During the first stage 
the molecule transitions from an extended random coil state to a compact state. The initial part of the collapse is characterized by the 
formation of short $3_{10}$-helices and a corresponding increase in asphericity and prolateness indicating a more rodlike structure. This is followed
by a decrease in $R_{\rm g}^2$ as well as the other shape factors to values corresponding roughly to a spherical state. Already during this process we 
observe formation of $3_{10}$- and $\alpha$-helices. The second stage of the process is characterized by a slight decrease in the number of $3_{10}$-helices
and the ongoing formation of new $\alpha$-helices and extension of existing $\alpha$-helices which eventually organize in equilibrium states such as 
helical bundles. 
Even though the shape parameters do not change during this process, the torsion energy appears to slightly increase. While a previous study\cite{Polyala}
found a separation between the formation of $\alpha$-helices and the collapse, we do not observe such a strict separation. A possible explanation could be
that we use chains that are significantly longer where equilibrium states would likely contain more than just one $\alpha$-helix.\cite{palencar_folding_2013} 

\subsection{Scaling laws governing the collapse of polyalanine}

\begin{figure}[t!]
    \centering
      \includegraphics[width=\linewidth]{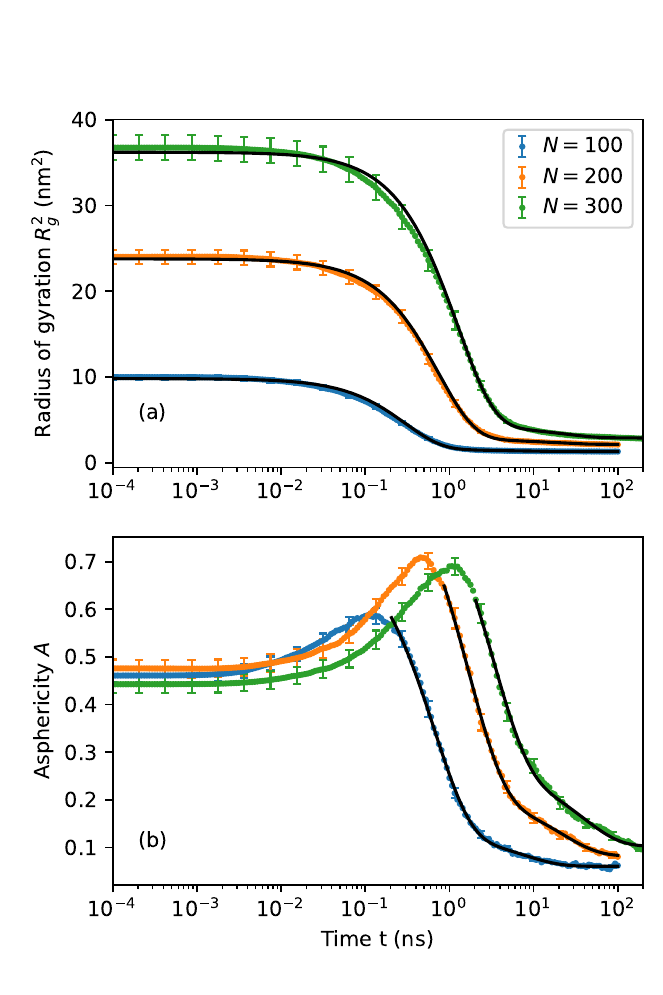} 
    \caption{Time evolution of (a) the squared radius of gyration and (b) the asphericity for different values of $N$ as indicated. All results 
    are averaged over the 200 (or 100 for $N\geqq125$) simulations with independent initial conformations. Legend for both frames is displayed in (a). 
    The black solid lines represent fits with the 2-exponential ansatz\ (\ref{2x}).}\label{A}
\end{figure}

Finally, by extracting relaxation times, we dive into a more quantitative analysis. To this end, we have added further simulations for chains of length 
$N = 25, 75 \dots 300$ with 200 realizations for $N\leqq100$ and 100 for $N>100$. All simulations are run up to at least $t = 100$ ns. In a first attempt 
to model the decay of $R_g^2$ from its random coil value at $t=0$ ($R_g^2\propto N^{1.2}$ where $1.2=2\nu_{SAW}$ with the Flory approximation 
$\nu_{SAW}=3/5$) to the close-to-asymptotic compact state ($R_g^2\propto  N^{2/3}$), we employed the template from studies of generic homopolymers\cite{maj:zie:jan,chr:maj:jan} 
and polyglycine\cite{maj:han:jan}, namely a stretched exponential (or Kohlrausch) ansatz
\begin{equation}
  R_{\text{g}}^2(t) = b_0 + b\mathrm{e}^{-{(t/\tau_{\text{c}})}^{\beta}}.
  \label{Rg2}
\end{equation}
This works fairly well also here and since the exponent $\beta$ turned out to be close to unity ($\beta\approx0.7$) even a simple exponential ansatz does 
qualitatively represent the decay, at least in its initial part. For the decay of the asphericity $A$ for times beyond its peak value, however, this is 
no longer true. Here, a much more satisfactory fit of the data for $t>t(A_{\rm max})$ is achieved with the 2-exponential form
\begin{equation}
  A(t) = a\mathrm{e}^{-t/\tau_{1}} + b\mathrm{e}^{-t/\tau_{2}} + c,
  \label{2x}
\end{equation}
where $\tau_1$ describes the initial decay (the overall ``collapse'') and $\tau_2\gg\tau_1$ the final, much slower relaxation process within the already 
formed compact state. Eventually, we used the same ansatz (\ref{2x}) for $R_g^2$ as well since also here it describes the data better than the stretched 
exponential (\ref{Rg2}) even though the difference is not very pronounced (we obtain $\tau_{\rm c}\approx\tau_1$ within a few percent). This good representation
can be seen from the fits in Fig.\ \ref{A} (a) for different chain lengths where the fits align with the data within error bars. Even better results were
obtained with this ansatz for the asphericity with $t>t(A_{\rm max})$ across different chain lengths shown in (b).

\begin{figure}[t!]
  \centering
  \includegraphics[width=\linewidth]{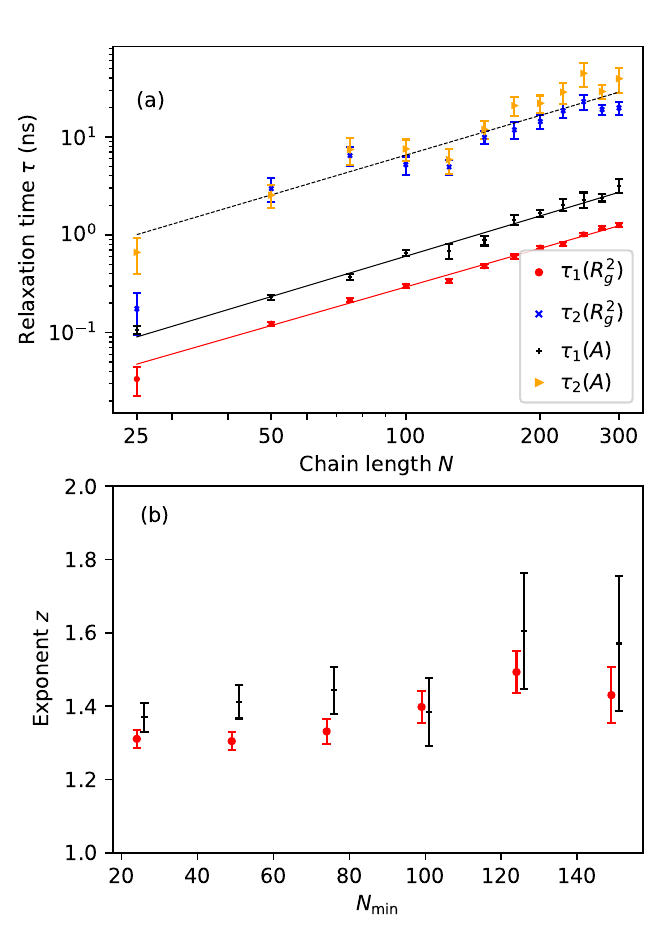}
\caption{(a) Relaxation times $\tau_1$ and $\tau_2$ obtained from ansatz (\ref{2x}) for different chain lengths $N$. The fitted lines correspond to 
$\tau\propto N^z$. The straight dotted line $\propto N^{1.35}$ through the data for $\tau_2$ is only a guide to the eye. All results are averaged over 
the 200 (or 100 for $N\geqq125$) simulations with independent initial conformations.
(b) Dependence of the scaling exponents $z$ on the minimum of the fitting range $N_{\rm \min}$ (for better visibility, the data for $A$ are displaced
to $N_{\rm min}+1$). Legend for both frames is displayed in (a).}\label{Lmin}
\end{figure}

Since the raw data of $R_g^2(t)$ and $A(t)$ are correlated in time we repeated all fits for delete-one Jackknife subsets\cite{efr,efr:ste,mil} and 
estimated the statistical errors on the fit parameters from the Jackknife variance. The resulting $\tau_1$ and $\tau_2$ are displayed in Fig.\ \ref{Lmin}. 
One can see in (a) that for both $R_g^2(t)$ and $A(t)$ the smaller relaxation times $\tau_1$ of the fast decay nicely follow the power law in $\tau_1\propto N^z$
represented by the straight lines. From the scaling of the relaxation time $\tau_1$ with $N$ of $R_g^2$ we obtain $z = 1.32(3)$, using the
fit over all data points (with $\chi^2/{\rm d.o.f.} = 2.3$), and for $A$ we find the compatible estimate $z = 1.38(4)$ (with $\chi^2/{\rm d.o.f.} = 1.2$).
These are the displayed fits. When changing the fit range, the estimates for $z$ vary only slightly as shown in (b). We hence conclude that the initial relaxation
(the overall ``collapse'') is characterized by $z \approx 1.35$.
The estimates for $\tau_2$ of the slower decay are less reliable since this final relaxation within the already formed compact structure probes the 
rightmost regime of $R_g^2(t)$ and $A(t)$ where the signal-to-noise ratio becomes rather poor. The estimated $\tau_2$ seem to also be compatible with a
power law $\tau_2\propto N^{1.35}$.
The result for the scaling exponent $z$ aligns with values typically found for homopolymers in MD simulations. It is, however, at odds with an earlier 
study on the collapse of polyglycine\cite{maj:han:jan} that found a very fast decay and conjectured that intrachain hydrogen bonds might be the reason why
biological polymers are able to fold so fast. A possible explanation for this difference is the formation of short helices early in the collapse
in our study. By locally stabilizing the polymer they seem to be related to the formation of a more rodlike structure during the collapse, which could be 
responsible for the slowing down. However, it would be interesting to investigate whether the use of explicit water could influence this behavior which is
our plan for future work. 

\section{Conclusion}
We have investigated the helix-coil transition in polyalanine using secondary structure analysis and shape factors derived from the gyration tensor. 
Based on the study of the collapse in molecules with a chain length of $N=100$, we observed that the transition in general can be split into two major 
stages: collapse and the subsequent transition into an equilibrium state with straight $\alpha$-helices. Already during the first stage helix formation 
is encountered, which may be responsible for the observed straightening of the molecule halfway through the collapse. 
The second stage of the transition is then characterized by the formation of straight $\alpha$-helices leading eventually to structures like helical 
bundles. 

In the third part of our analysis, we applied a 2-exponential ansatz to the time evolutions of radius of gyration and the asphericity. For both observables
we find that the smaller relaxation time $\tau_1$ scales with an exponent of $z\approx1.35$, similar to exponents obtained in previous studies for homopolymers.  
The larger relaxation times $\tau_2$ also appear to follow a power law with exponent $z\approx1.35$. It would be interesting to investigate the
scaling of this relaxation time with longer simulations to better capture this relaxation on larger time scales.

Our results suggest that the collapse of polyalanine is influenced by the formation of $3_{10}$-helices and $\alpha$-helices. The formation of these 
structures during the collapse could slow down this process compared to other polymers. An interesting extension of this study would be inclusion of explicit solvent which may 
allow to better disentangle the different factors influencing the two relaxation times.

\begin{acknowledgments}
  This project was funded by the Deutsche Forschungsgemeinschaft (DFG, German Research Foundation) – 469 830 597 under project ID JA 483/35--1. S.M.
  thanks the Science and Engineering Research Board (SERB), Govt.\ of India
  for a Ramanujan Fellowship (file no. RJF/2021/000044). All simulations were
  performed on the GPU cluster of the Universitätsrechenzentrum (URZ) at Universität Leipzig.
\end{acknowledgments}

\section*{Author Declarations}

\subsection*{Conflict of interest}

The authors have no conflicts to disclose.

\section*{Data Availability Statement}

Data available on request from the authors.

\section*{References}
\bibliography{references_new}

\end{document}